\documentclass[12pt,a4paper]{article}

\usepackage{epsfig}

\begin{document}

\def\lp{\left. }
\def\rp{\right. }
\def\lr{\left( }
\def\rr{\right) }
\def\le{\left[ }
\def\re{\right] }
\def\lg{\left\{ }
\def\rg{\right\} }
\def\lb{\left| }
\def\rb{\right| }

\def\go{\tilde{g}}
\def\mg{m_{\go}}

\def\msQ {m_{\tilde{Q}}}
\def\msD {m_{\tilde{D}}}
\def\msU {m_{\tilde{U}}}
\def\msS {m_{\tilde{S}}}
\def\msC {m_{\tilde{C}}}
\def\msB {m_{\tilde{B}}}
\def\msT {m_{\tilde{T}}}
\def\msusy{m_{\rm SUSY}}

\def\sq  {\tilde{q}}
\def\sql {\tilde{q}_L}
\def\sqr {\tilde{q}_R}
\def\ms  {m_{\sq}}
\def\msql{m_{\tilde{q}_L}}
\def\msqr{m_{\tilde{q}_R}}

\def\st  {\tilde{t}}
\def\stl {\tilde{t}_L}
\def\str {\tilde{t}_R}
\def\mstl{m_{\stl}}
\def\mstr{m_{\str}}
\def\sta {\tilde{t}_1}
\def\stb {\tilde{t}_2}
\def\msta{m_{\sta}}
\def\mstb{m_{\stb}}
\def\thst{\theta_{\tilde{t}}}

\def\sb  {\tilde{b}}
\def\sbl {\tilde{b}_L}
\def\sbr {\tilde{b}_R}
\def\msbl{m_{\sbl}}
\def\msbr{m_{\sbr}}
\def\sba {\tilde{b}_1}
\def\sbb {\tilde{b}_2}
\def\msba{m_{\sba}}
\def\msbb{m_{\sbb}}
\def\thsb{\theta_{\tilde{b}}}

\newcommand{\sfa}{\tilde{f}}

\newcommand{\SLASH}[2]{\makebox[#2ex][l]{$#1$}/}
\newcommand{\kslash}{\SLASH{k}{.15}}
\newcommand{\pslash}{\SLASH{p}{.2}}
\newcommand{\qslash}{\SLASH{q}{.08}}

\renewcommand{\thefootnote}{\fnsymbol{footnote}}

\def\d  {{\rm d}}
\def\eps{\varepsilon}

\def\beq{\begin{equation}}
\def\eeq{\end{equation}}
\def\bea{\begin{eqnarray}}
\def\eea{\end{eqnarray}}

\begin{titlepage}
\hfill hep-ph/0210420 \\ \mbox{}\hfill October 2002 \\[1cm]

\begin{center}
{\large\bf The Production of Gluino Pairs in High Energy \boldmath $e^+e^-$
Collisions}%
\footnote{To appear in the Proceedings of the 10$^{\rm th}$
International Conference on Supersymmetry and Unification of Fundamental
Interactions (SUSY02), June 17-23, 2002, DESY Hamburg.}\
\\[1.5cm]

{\sc Stefan BERGE and Michael KLASEN}\footnote{michael.klasen@desy.de} \\[5mm]

II.\ Institut f\"ur Theoretische Physik, Universit\"at Hamburg, \\
Luruper Chaussee 149, D-22761 Hamburg, Germany \\[1.5cm]
\end{center} 

\begin{abstract}
In the Minimal Supersymmetric Standard Model (MSSM), the process
$e^+e^-\to\go\go$ is mediated by
quark/squark loops, dominantly of the third generation, where the mixing of
left- and right-handed states can become large. Taking into account realistic
beam polarization effects, photon and $Z^0$-boson exchange, and current mass
exclusion limits, we scan the MSSM parameter space for various $e^+e^-$
center-of-mass energies to determine the regions, where gluino production
should be visible.
\end{abstract}

\end{titlepage}


\section{Introduction}
\label{sec:1}

In the Minimal Supersymmetric Standard Model (MSSM) \cite{Nilles:1983ge,
Haber:1984rc}, the exclusive production of gluino pairs in $e^+e^-$
annihilation is mediated by $s$-channel photons and $Z^0$-bosons,
which couple to the gluinos via triangular quark and squark loops.
In this Talk, we report about a recent study \cite{Berge:2002ev}
of the potential of high-energy
linear $e^+e^-$ colliders for the production of gluino pairs within the MSSM.
Similar studies have been performed earlier for the production of sfermion
pairs \cite{Berge:2000cb,Klasen:2000cc}.
Taking into account realistic beam polarization effects, photon and $Z^0$-boson
exchange, and current mass exclusion limits, we scan the MSSM parameter space
for various $e^+e^-$ center-of-mass energies to determine the regions, where
gluino production should be visible. Furthermore we clarify the theoretical
questions of the relative sign between the two contributing triangular Feynman
diagrams, of the possible presence of an axial vector anomaly, and the
conditions for vanishing cross sections -- three related issues, which have so
far been under debate in the literature.

\section{Results}
\label{sec:2} 

\begin{figure}
 \centering
 \epsfig{file=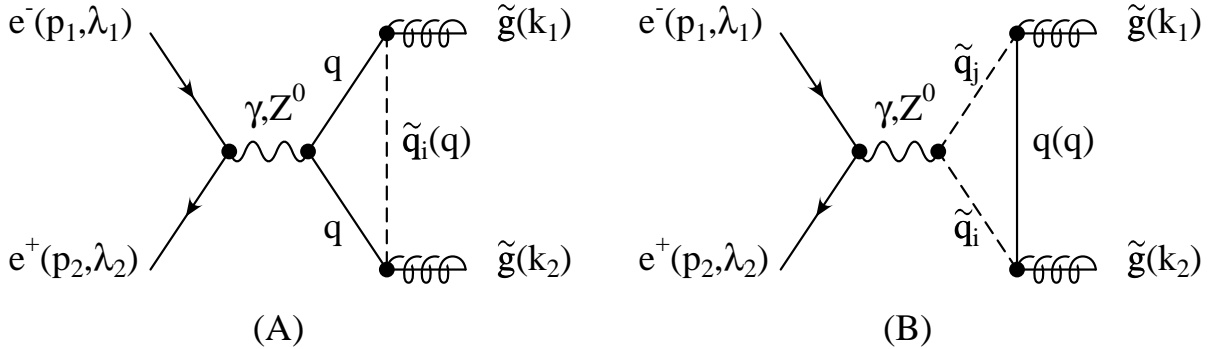,bbllx=38pt,bblly=349pt,bburx=314pt,bbury=431pt,%
  width=\columnwidth}
 \caption{\label{klafig:1}Feynman diagrams for gluino pair production in
 electron-positron annihilation. The exchanged photons and $Z^0$-bosons
 couple to the produced gluinos through triangular $qq\sq_i$ (A) and $\sq_i
\sq_j q$ (B) loops with flavor flow in both directions.}
\end{figure}

The scattering process
\beq
 e^-(p_1,\lambda_1)\ e^+(p_2,\lambda_2)\to\go(k_1)\ \go(k_2)
\eeq
with incoming electron/positron momenta $p_{1,2}$ and helicities
$\lambda_{1,2}$ and outgoing gluino momenta $k_{1,2}$ proceeds through
the two Feynman diagrams A and B in Fig.\ \ref{klafig:1}
with $s$-channel photon and $Z^0$-boson exchange and triangular quark and
squark loops. Higgs boson exchange is not considered due to the negligibly
small electron Yukawa coupling, but it could well be relevant at
muon colliders. The process occurs only at the one-loop level, since the gluino
as the superpartner of the gauge boson of the strong interaction couples
neither directly to leptons nor to electroweak gauge bosons. Our analytical
results are presented in Ref.\ \cite{Berge:2002ev} in full detail. Taking into
account chiral squark mixing, we find that pairs of identical (Majorana)
gluinos are produced by a parity violating axial vector coupling induced by
mass differences between the chiral squarks and the axial vector coupling of
the $Z^0$-boson. The (mass-independent) ultraviolet singularities contained in
the loop integrals cancel among the two contributing Feynman diagrams.
As we have checked explicitly (even for complex squark mixing matrices),
adding the two amplitudes induces not only a cancellation of the ultraviolet
singularities and of the logarithmic dependence on the scale parameter, but
also a destructive interference of the finite remainders. This happens
separately for each weak isospin partner. Our analytical results have been
obtained in two independent analytical calculations and have been implemented
in a compact Fortran computer code. As a third independent cross-check, we have
recalculated the production of gluino pairs in $e^+e^-$ annihilation with the
computer algebra program FeynArts/FormCalc \cite{Hahn:2001rv} and found
numerical agreement up to 15 digits.

Our calculations involve various masses and couplings of SM particles,
for which we use the most up-to-date values from the 2002 Review of
the Particle Data Group \cite{Hagiwara:pw}. In particular, we evaluate the
electromagnetic fine structure constant $\alpha(m_Z)=1/127.934$ at the mass of
the $Z^0$-boson, $m_Z = 91.1876$ GeV, and calculate the weak mixing angle
$\theta_W$ from the tree-level expression $\sin^2\theta_W=1-m_W^2/m_Z^2$ with
$m_W = 80.423 $ GeV. Among the fermion masses, only the one of the top quark,
$m_t = 174.3$ GeV, plays a significant role due to its large splitting from the
bottom quark mass, $m_b = 4.7$ GeV, while the latter and the charm quark mass,
$m_c = 1.5$ GeV, could have been neglected like those of the three light quarks
and of the electron/positron. The strong coupling constant is evaluated at the
gluino mass scale from the one-loop expression with five active flavors and
$\Lambda_{\rm LO}^{n_f=5} = 83.76$ MeV, corresponding to $\alpha_s(m_Z) =
0.1172$. A variation of the renormalization scale by a factor of four about the
gluino mass results in a cross section uncertainty of about $\pm25$ \%. Like
the heavy top quark, all SUSY particles have been decoupled from the running
of the strong coupling constant.

%
%

We work in the framework of the MSSM with conserved $R$- (matter-) parity,
which represents the simplest phenomenologically viable model, but which is
still sufficiently general to not depend on a specific SUSY breaking mechanism.
Models with broken $R$-parity are severely restricted by the non-observation
of proton decay, which would violate both baryon and lepton number
conservation.
We do not consider light gluino mass windows, on which the literature has
focused so far and which may or may not be excluded from searches at fixed
target and collider experiments \cite{Hagiwara:pw}. Instead, we adopt the
current mass limit $\mg\geq 200$ GeV from the CDF \cite{Affolder:2001tc} and
D0 \cite{Abachi:1995ng} searches in the jets with missing energy channel,
relevant for non-mixing squark masses of $\ms\geq 325$ GeV and $\tan\beta = 3$.
Values for the ratio of the Higgs vacuum expectation values, $\tan\beta$, below
2.4 are already excluded by the CERN LEP experiments, although this value is
obtained using one-loop corrections only and depends in addition on the top
quark mass \cite{lhwg:2001xx}. In order to delimit the regions of large
gluino cross sections, we have scanned the MSSM parameter space over
$\tan\beta\in[1.6;50]$, the Higgs mass parameter $\mu=[-2;2]$ TeV, the
trilinear coupling $A_q\in[-6;6]$ TeV, and the squark mass parameter
$\msusy\in[200;2000]$ GeV. We found that the cross section is visible only
for parameter choices resulting in large squark mass splittings, specified
below, and that its sensitivity to individual parameter choices is small.

%
\begin{figure}
 \centering
 \epsfig{file=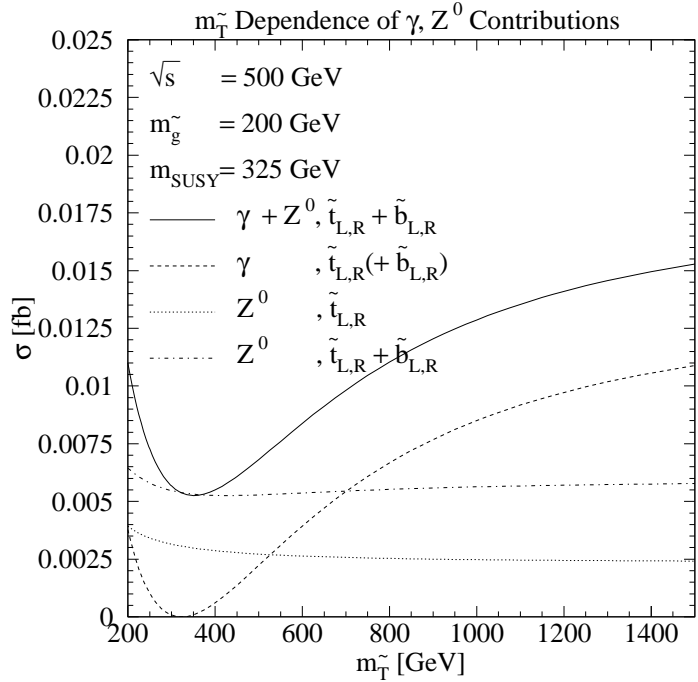,width=0.6\columnwidth}
 \caption{\label{klafig:4}Dependence of the photon and $Z^0$-boson contributions
 to the process $e^+e^-\to\go\go$ on the right-handed top squark mass parameter
 $\msT$. The photon contribution (dashed curve) is dominated by top (s)quarks
 and cancels for $\mstl=\mstr$. 
}
\end{figure}
%

If not stated otherwise, we will present
unpolarized cross sections for a $\sqrt{s}=500$ GeV
linear $e^+e^-$ collider like DESY TESLA, gluino masses of $\mg=200$ GeV, and
squark masses $\ms\simeq\msQ=\msD=\msU=\msS=\msC=\msB=\msT=\msusy=325$ GeV.
We will consider two cases of large squark mass splittings: I.) On the one
hand, the masses of the superpartners of left- and right-handed
quarks need not be equal to each other. In this scenario we will vary the
right-handed up-type squark mass parameters
$m_{\tilde{U},\tilde{C},\tilde{T}}$ between 200 and 1500 GeV. II.) On the
other hand, the superpartners of the heavy quarks can mix into light and heavy
mass eigenstates. This alternative is restricted by
the CERN LEP limits on the light top and bottom squark masses, $\msta\geq 100$
GeV and $\msba\geq 99$ GeV \cite{lswg:2002xx}, and on SUSY one-loop
contributions \cite{Barbieri:1983wy,Drees:1990dx,Chankowski:1993eu} to the
$\rho$-parameter, $\rho_{\rm SUSY} < 0.0012$ \cite{Hagiwara:pw}.
In this case we assume the maximally
allowed top squark mixing with $\thst=45.2^\circ$, $\msta=110$ GeV, and
$\mstb=506$ GeV, which can be generated by choosing appropriate values for the
Higgs mass parameter, $\mu=-500$ GeV, and the trilinear top squark coupling,
$A_t=534$ GeV. For small values of $\tan\beta$, mixing in the bottom squark
sector remains small, and we take $\theta_{\sb}=0^\circ$.
Although the absolute magnitude of the cross section depends strongly on the
gluino mass and collider energy, the relative importance of the different
contributions is very similar also for higher gluino masses and collider
energies.

First we examine the conditions found in Ref.\ \cite{Berge:2002ev} for
vanishing of
the photon and $Z^0$-boson contributions, restricting ourselves to the
third generation. Since we expect the photon contribution to cancel for equal
left- and right-handed squark masses, we vary the right-handed top squark mass
parameter, $\msT\simeq\mstr$, between 200 and 1500 GeV, but keep $\mstl\simeq
\msQ=\msB=\msusy=325$ GeV fixed (case I), since top and bottom squarks
generally interfere destructively due to their opposite charge and weak isospin
quantum numbers. As can be seen from Fig.\
\ref{klafig:4}, the photon contribution cancels indeed for
$\msT\simeq\mstr=\mstl\simeq\msusy$, {\it i.e.} for all flavors $q$
with equal squark masses. Due
to their charge, top (s)quarks contribute four times as much as bottom
(s)quarks, whose contribution is even more suppressed by the condition $\msbl
\simeq\msbr$. The $Z^0$-boson contribution can never cancel, since $m_t\gg
m_b$, and therefore it depends only weakly on $\msT$, but it can become minimal
for $\msT\simeq\mstr=\mstl=\msbr=\msbl\simeq\msusy$. As $\msT$ gets
significantly larger (or smaller) than $\msusy$, the photon contribution
starts to dominate over the $Z^0$-boson contribution.
If only $\msT$ differs from $\msusy$, the third generation contributes almost
100\% to the total cross section. However, if $\msU=\msC=\msT$ are varied
simultaneously, all three generations contribute to the total cross section,
which can therefore become significantly larger.

\begin{figure}
 \centering
 \epsfig{file=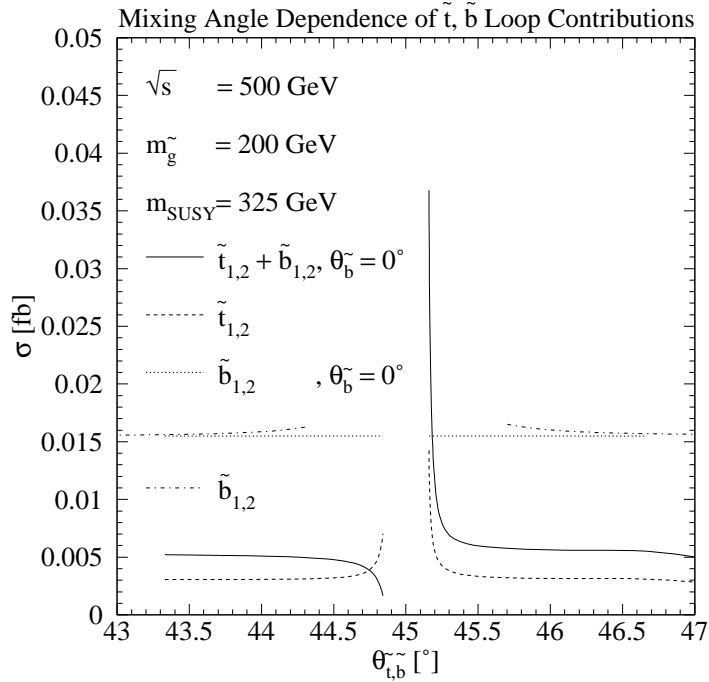,width=0.6\columnwidth}
 \caption{\label{klafig:6}Mixing angle dependence of the $\st$ (dashed) and $\sb$
 (dotted) loop contributions to the process $e^+e^-\to\go\go$, which interfere
 destructively (full curve), except for $\thst\simeq 45.2^\circ$, where the
 imaginary parts of the amplitudes interfere constructively. Mixing in the
 $\sb$ sector (dot-dashed curve) enhances the cross section only slightly.}
\end{figure}

When $\msU=\msC=\msT=\msusy$ and large mass splittings are generated only by
mixing in the top squark sector (case II), photon contributions are suppressed
by more than two orders of magnitude. Fig.\ \ref{klafig:6} shows that the
%
%
$Z^0$-boson contributions from top and bottom squarks interfere destructively
due to opposite values of their weak isospin quantum numbers, except for
$\theta_{\st}\simeq 45.2^\circ$, where the imaginary parts of the amplitudes
interfere constructively. It is therefore advantageous to keep the bottom
squark mass splitting small. As is also evident from Fig.\ \ref{klafig:6}, mixing
in the bottom squark sector is of little importance. Note that
the central region with maximal top/bottom squark mixing is excluded by the
CERN LEP limits on $\msta$, $\msba$, and the $\rho$-parameter.
When $\msusy$ and the diagonal elements of the squark mixing matrix
become much larger than the quark masses and the off-diagonal
elements of the matrix, the role of squark mixing is reduced, as expected.
Squarks from the first two generations
contribute at most 10\% at low $\msusy$ and are otherwise strongly suppressed.

\begin{figure}
 \centering
 \epsfig{file=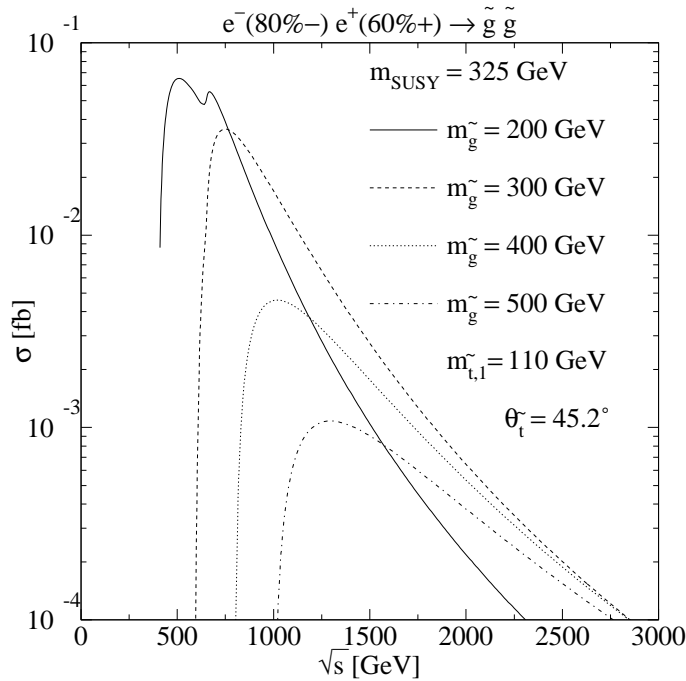,width=0.6\columnwidth}
 \caption{\label{klafig:9}Center-of-mass energy dependence of the polarized
 $e^+e^-\to\go\go$ cross section for various gluino masses and maximal
 top squark mixing.}
\end{figure}

At future linear $e^+e^-$ colliders it will be possible to obtain relatively
high degrees of polarization, {\it i.e.} about 80\% for electrons and 60\% for
positrons.
For these degrees of polarization, we show in
Fig.\ \ref{klafig:9} a scan in the center-of-mass energy of a future $e^+e^-$
%
%
collider for various gluino masses and maximal top squark mixing (case II).
The cross section rises rather slowly due to the factor $\beta^3$
for $P$-wave production of the gluino pairs. For $\mg=200$
GeV we observe an interesting second maximum, which arises from the
intermediate squark pair resonance at $\sqrt{s}=2\,\msusy=650$ GeV. At
threshold, the cross section depends strongly on the gluino mass and is largest
for $\mg= 200$ GeV, which we consider to be the lowest experimentally allowed
value. It drops fast with $\mg$, so that for $\mg>500$ GeV no events at
colliders with luminosities of 1000 fb$^{-1}$ per year can be expected,
irrespective of their energy. Smaller squark mixing (cf.\ Fig.\ \ref{klafig:6})
or larger values of $\msusy$ will reduce the cross
section even further. Far above threshold, it drops off like $1/s$ and
becomes independent of the gluino mass.

\begin{figure}
 \centering
 \epsfig{file=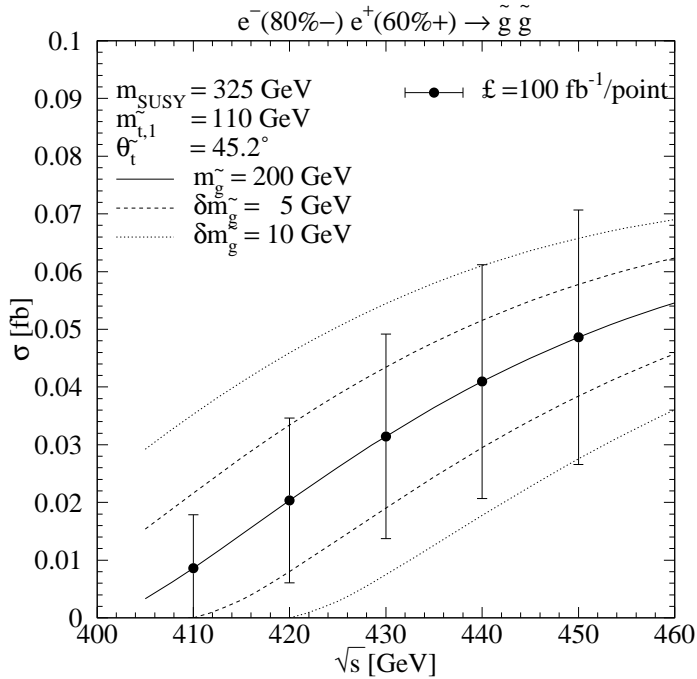,width=0.6\columnwidth}
 \caption{\label{klafig:11}Sensitivity of the polarized $e^+e^-\to\go\go$ cross
 section to the gluino mass $\mg$ for maximal top squark mixing. The central
 values and statistical error bars of the data points have been calculated
 assuming $\mg=200$ GeV and a luminosity of 100 fb$^{-1}$ per center-of-mass
 energy point.}
\end{figure}

The slow rise of the cross section can be observed even better in Fig.\
\ref{klafig:11}, where the sensitivity of a $\sqrt{s}=500$ GeV collider like
%
%
DESY TESLA to gluino masses around 200 GeV has been plotted. For the CERN LHC
experiments, a precision of $\pm30\,...\,60$ ($12\,...\,25$) GeV is expected
for gluino masses of 540 (1004) GeV. If the
masses and mixing angle(s) of the top (and bottom) squarks are known, a
precision of $\pm5\,...\,10$ GeV can be achieved at DESY TESLA for $\mg=200$
GeV and maximal top squark mixing with an integrated luminosity of 100
fb$^{-1}$ per center-of-mass energy point.

As has already been mentioned above, the Majorana nature of the gluino leads
to a vanishing forward-backward asymmetry. In order to establish this feature
experimentally, the asymmetry has to be measured with an accuracy of at least
7--10\%, which are the values relevant for Dirac fermions such as up-
(charm) and down-type (strange/bottom) quarks \cite{Hagiwara:pw}. As a
consequence, the production cross section must be known with about the same
precision. In view of the results obtained in Fig.\ \ref{klafig:11}, such a
measurement appears to be extremely difficult if not impossible.

\section{Conclusions}
\label{sec:3}

In this Talk, we have reported on a recent study of gluino pair production at
future $e^+e^-$ colliders \cite{Berge:2002ev}. In this study, we have resolved
a long-standing discrepancy in the literature about the relative sign of the
quark and squark loop contributions to the production of gluino pairs in
$e^+e^-$ annihilation. We have found that the ultraviolet divergence cancels
for each squark flavor separately and not between weak isospin partners. Our
results rely on two completely independent analytical calculations and one
computer algebra calculation.
Furthermore, we have investigated the prospects for precision measurements of
gluino properties, such as its mass or its Majorana fermion nature, at future
linear $e^+e^-$ colliders. We have taken into account realistic beam
polarization effects, photon and $Z^0$-boson exchange, and current mass
exclusion limits. Previously, only light gluinos at center-of-mass energies up
to the $Z^0$-boson mass had been investigated. Within the general framework of
the MSSM, we have concentrated on two scenarios of large left-/right-handed
up-type squark mass splitting and large top squark mixing, which produce
promisingly large cross sections for gluino masses up to 500 GeV or even
1 TeV. Gluino masses of 200 GeV can then be measured with a precision of about 
5 GeV in center-of-mass energy scans with luminosities of 100 fb$^{-1}$/point.
However, when both the left-/right-handed squark mass splitting and the
squark mixing remain small, gluino pair production in $e^+e^-$ annihilation
will be hard to observe, even with luminosities of 1000 fb$^{-1}$/year.


\section*{Acknowledgments}

This work has been supported by Deutsche Forschungsgemeinschaft through
Grant No.\ KL~1266/1-2 and Graduiertenkolleg {\it Zuk\"unftige
Entwicklungen in der Teilchenphysik}.



\end{document}